\documentstyle[psfig]{europhys} 

\def\etal{{\hbox{{\tenit\ et al.\/}\tenrm :\ }}}

\def\And{{\rm and\ }}

\def\stars{\bigskip\centerline{***}\medskip}

\newif\ifboo \boofalse

\def\Review#1{\boofalse{\it #1},}
\def\Name#1{{\sc #1},}
\def\Vol#1{\ifboo Vol. {\bf #1}\else{\bf #1}\fi}
\def\Year#1{\ifboo #1\else(#1)\fi}
\def\Book#1{\bootrue{\it #1},}
\def\Page#1{\ifboo {\rm p. #1}\else{\rm #1}\fi}

\begin{document}
\euro{xx}{x}{xx-xx}{xxxx}
\Date{}
\shorttitle{C. GRIMALDI \etal ISOTOPE EFFECT ON $m^*$
ETC.}
\title{{\Large Isotope Effect on $m^{*}$ in High-$T_{c}$ Materials
due to the Breakdown of Migdal's Theorem}} 
\author{C. Grimaldi\inst{1}, E. Cappelluti\inst{2}, and
L. Pietronero\inst{1}\inst{3}}
\institute{\inst{1} Dipartimento di Fisica, Universit\'{a} di Roma ``La Sapienza", 
Piazzale A.  Moro, 2, 00185 Roma, Italy 
and Istituto Nazionale Fisica della Materia, Sezione di Roma 1, Italy \\
\inst{2} Max-Planck-Institut f\"{u}r Festk\"{o}rperforschung, 
Heisenbergstrasse  1, D-70569 Stuttgart, Germany\\
\inst{3} ICTP, P.O.  Box 586, 34100 Trieste, Italy}
\rec{}{}
\pacs{
\Pacs{71}{38$+i$}{Polarons and electron-phonon interactions}
 \Pacs{74}{25Kc}{Phonons}
\Pacs{71}{18$+y$}{Fermi surface: 
calculations and measurements; effective mass, g factor}
}

\maketitle
\begin{abstract}
We show that the inclusion of effects beyond Migdal's limit
in the electron-phonon interaction naturally leads to an
isotope effect for the effective mass $m^{*}$ of the charge carriers
even much before reaching the polaron limit.
This is the situation already considered in our approach to 
nonadiabatic superconductivity \cite{grima1}.
Such a result provides a scenario different from the polaronic one 
for the interpretation of the recently observed isotope effect on $m^{*}$ in  
YBa$_{2}$Cu$_{3}$O$_{6+x}$ and La$_{2-x}$Sr$_{x}$CuO$_{4}$ \cite{zhao1,zhao2}.
\end{abstract}

The recent observation of an oxygen-mass dependent penetration depth 
$\lambda(0)$ in YBa$_{2}$Cu$_{3}$O$_{6+x}$ \cite{zhao1}
and in La$_{2-x}$Sr$_{x}$CuO$_{4}$ \cite{zhao2} has raised the
question whether this effect could be attributed to the breakdown of
the Migdal approximation \cite{migdal}.
In fact, according to the classical Migdal-Eliashberg theory of 
superconductivity, the zero temperature penetration depth in the 
London limit and in the absence of magnetic impurities \cite{kresin} is
given by:

\begin{equation}
\label{depth}
\lambda(0)\propto \sqrt{\frac{m^{*}}{n_{s}}} ,
\end{equation}
where $m^{*}$ is the electronic effective mass and $n_{s}$ is the 
supercarrier density. 
According to eq.(\ref{depth}),
an isotope effect on $\lambda(0)$ can therefore be induced 
by an isotope effect on $m^{*}$ and/or on $n_{s}$. 
However, for $T\ll T_{c}$, also assuming an electron-phonon (el-ph)
type of pairing, the dependence of
$n_{s}$ on the ion-mass $M$ is negligible \cite{abri} and a 
possible isotope-induced dependence of 
$n_{s}$ on the hole concentration has been ruled out \cite{zhao1,zhao2}.
Zhao et al. therefore concluded
that the observed isotope effect on $\lambda(0)$ can be 
entirely attributed to an oxygen-mass dependent $m^{*}$.
In this hypothesis and by using eq.(\ref{depth}), the oxygen isotope effect on $m^{*}$, 
$\alpha_{m^{*}}=-d\ln(m^{*})/d\ln(M)$,
has been estimated to be $\alpha_{m^{*}}=-0.61\pm 0.09$ for
YBa$_{2}$Cu$_{3}$O$_{6.94}$ \cite{zhao1} and $\alpha_{m^{*}}\simeq -0.8\,(-0.5)$ for 
La$_{2-x}$Sr$_{x}$CuO$_{4}$ for $x=0.105\,(0.15)$ \cite{zhao2}. 
These large negative 
values of $\alpha_{m^{*}}$ cannot be explained by the classical 
Migdal-Eliashberg (ME) theory of the el-ph interaction, which 
states that $\alpha_{m^{*}}=0$.
In fact this theory predicts that the electron-mass renormalization factor $m^{*}/m=Z_{0}$ 
is given by $Z_{0}=1+\lambda$, where $\lambda$ is the el-ph coupling constant, 
{\it i.e.}, a quantity independent of the ion-mass $M$
 ($\lambda$ must not be confused 
with $\lambda(0)$, the penetration depth). 
This is a consequence of the adiabatic condition 
$\lambda\,\omega_{D}/(q v_{F})\ll 1$,
where $\omega_{D}$ is the Debye phonon frequency, 
$v_{F}$ is the Fermi velocity and
$q$ is the momentum transfer in the el-ph scattering. The adiabatic condition
permits to neglect the vertex corrections to the el-ph interaction 
(Migdal approximation) and represents the basis
of the formulation of the ME theory. 
The unusual isotope effect $\alpha_{m^{*}}$ observed in \mbox{} YBa$_{2}$Cu$_{3}$O$_{6+x}$ and in 
La$_{2-x}$Sr$_{x}$CuO$_{4}$ could therefore be interpreted 
as a clear signature of the breakdown of the Migdal approximation.

However, from the quantitative point of view,
some care has to be used in estimating $\alpha_{m^{*}}$ from the isotope effect of 
$\lambda(0)$ via eq.(\ref{depth}), since the latter applies only when the Migdal 
approximation is a valid assumption. A generalization of eq.(\ref{depth}) beyond 
Migdal's theorem requires in fact the knowledge of the Eliashberg equations for 
temperatures below $T_{c}$ with the inclusion of the nonadiabatic contributions, 
{\it i.e.}, a task which even in the perturbative approach presents serious technical
difficulties. 
Nevertheless, the measurements reported in Refs.\cite{zhao1,zhao2}  
lead to two important consequences. First, the observation of an isotope effect
on $\lambda(0)$ implies a significant role of the el-ph interaction in the
high-$T_{c}$ compounds. Second, as pointed out before, such an the el-ph
interaction falls outside of the validity of the Migdal approximation.
Therefore, the findings of Refs.\cite{zhao1,zhao2} open new perspectives
in investigating unusual isotope effects on quantities which, according to the classical 
(in the sense of ME) theory of the el-ph interaction, 
should not show isotope effects at all. 

One of the most unambiguous evidences for the breakdown of the ME theory
could be provided by the observation {\it in the
normal state} of an isotope effect on the effective electronic mass $m^{*}$.
In principle, isotope sensitive specific heat measurements could be able
to observe such an effect provided that the electronic contribution to
the specific heat can be clearly singled out \cite{loram}. 
With this motivation, we consider in this paper the consequences of the breakdown
of Migdal's theorem on $m^{*}$ and estimate 
the coefficient $\alpha_{m^{*}}$ of the electron-mass isotope effect. We show that 
nonzero values of $\alpha_{m^{*}}$ in the normal state can be obtained by considering 
two different regimes of the el-ph coupled system:
the polaronic state and the ``nonadiabatic regime" as described by the inclusion
of the first el-ph vertex correction in the electronic
self-energy. We have already studied the latter situation in connection with the 
superconductive transition \cite{grima1,grima2} and 
here we show that this theory, which does not imply the crossover towards the
polaronic state, naturally leads to an isotope effect for $m^{*}$.

Let us start our discussion with the polaron model. A similar analysis has already 
been performed in Ref.\cite{zhao2} in connection with the experimentally observed 
isotope effect on the penetration depth $\lambda(0)$
\cite{zhao1,zhao2}. For simplicity, we consider the Holstein model \cite{holstein} 
where phonons with frequency $\omega_{0}$ are locally coupled to electrons through 
a structureless el-ph coupling $g$ (small polaron). 
A polaronic state is characterized by a strong
electron-lattice correlation \cite{capone} and/or a large effective mass \cite{ciuchi}.
Within the range of applicability of the Holstein approximation, the effective
polaron mass is given as follows:

\begin{equation}
\label{pola1}
m^{*}\simeq m\exp\left(\frac{g^{2}}{\omega_{0}^{2}}\right),
\end{equation}
where $m$ is the bare electron mass.
Equation (\ref{pola1}) applies to the el-ph system in the antiadiabatic
regime $\omega_{0}/E_{F} > 1$, provided that the quantity $g^{2}/\omega_{0}$
is not too large \cite{ciuchi}.
Since $g^{2}\propto (M\omega_{0})^{-1}$ and $\omega_{0}^{2}\propto 1/M$,
from eq.(\ref{pola1}) we obtain that $m^{*}\propto\exp\sqrt{\gamma M}$, 
where $\gamma$ is a constant independent of $M$. 
The isotope effect on $m^{*}$ is therefore given by:

\begin{equation}
\label{pola2}
\alpha_{m^{*}}=-\frac{d\ln(m^{*})}{d\ln(M)}=-
\frac{1}{2}\left(\frac{g^{2}}{\omega_{0}^{2}}\right) .
\end{equation}
The above expression gives a negative value of $\alpha_{m^{*}}$ in
accordance therefore with the results reported in Refs.\cite{zhao1,zhao2}.
We must stress however that this result is based on eq.(\ref{pola1}), 
which holds true only as long as $\omega_{0}/E_{F} > 1$, {\it i. e.},
a rather inadeguate limit  for  YBa$_{2}$Cu$_{3}$O$_{6+x}$ 
and La$_{2-x}$Sr$_{x}$CuO$_{4}$ \cite{plakida}.
It is interesting to notice that, at optimal doping, LSCO shows 
a large negative $\alpha_{m^{*}}$ \cite{zhao2} and a negligible
isotope coefficient of the critical temperature $\alpha_{T_c}$ \cite{franck}.  
This behavior is therefore in contrast with the prediction of the bi-polaronic
theory of superconductivity which claims that 
$\alpha_{m^{*}}\propto \alpha_{T_c}$.

\begin{figure}
\vbox to 4cm{\vfill
\centerline{\psfig{figure=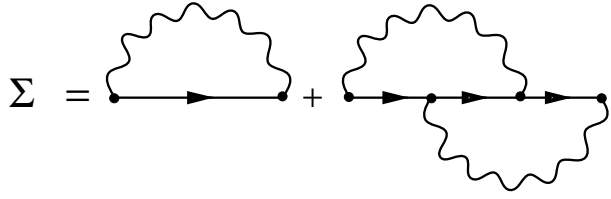,width=8cm}}\vfill}
\caption{Electronic self-energy beyond Migdal's approximation. Solid lines
represent electronic Green's functions while the wavy lines are
phonon propagators.}
\label{selfen}
\end{figure}

Besides the ME and polaronic scenarios, the el-ph coupled system
may display also a regime which is beyond Migdal's limit 
(and therefore beyond ME framework) and well
separated from the crossover between the quasi free-electron 
and the polaronic state.
Such a regime, which we call nonadiabatic,
is characterized by quasi free-electron states ($\lambda<\sim1$)
coupled in a nonadiabatic way to the lattice so that the el-ph
vertex corrections are relevant. 
In the nonadiabatic regime,
the electronic self-energy $\Sigma(i\omega_{n})$ is 
given by the graph depicted in Fig. \ref{selfen}
where the first vertex correction has been included.
The mass renormalization factor $Z_{0}=m^*/m$ is obtained by the 
$i\omega_n \rightarrow 0$ limit of
$Z(i\omega_{n})=1-\Sigma(i\omega_{n})/(i\omega_{n})$. The function
$Z(i\omega_n)$ has been obtained 
in a previous work \cite{grima2} and its expression is reported below:

\begin{equation}
\label{migdal1}
Z(i\omega_{n})=  1  +\frac{\pi T}{\omega_{n}}\sum_{m}
\frac{\lambda_{Z}(i\omega_{n},i\omega_{m};Q_{c})\omega_{0}^{2}}
{(\omega_{n}-\omega_{m})^{2}+\omega_{0}^{2}}
\frac{\omega_{m}}{|\omega_{m}|} \frac{2}{\pi}
\arctan\left(\frac{E/2}{|\omega_{m}|Z(i\omega_{m})}\right) .
\end{equation}
Here, $\omega_{n}$ and $\omega_{m}$ are fermionic Matsubara frequencies
and $\lambda_{Z}(i\omega_{n},i\omega_{m};Q_{c})$ is the frequency dependent 
el-ph coupling resulting from the inclusion of the first vertex correction 
function $P_{V}(i\omega_{n},i\omega_{m};Q_{c})$ into the electronic self-energy:

\begin{equation}
\label{migdal2}
\lambda_{Z}(i\omega_{n},i\omega_{m};Q_{c})=\lambda
\left[1+\lambda P_{V}(i\omega_{n},i\omega_{m};Q_{c})\right] .
\end{equation}
The explicit expression of the vertex function $P_{V}$ has already been presented
in Ref.\cite{grima2} for the three dimensional case and in Ref.\cite{baff1} for
the two dimensional one. The dimensionless parameter $Q_{c}=q_{c}/(2k_{F})$,
where $q_{c}$ is a cut-off over the momentum transfer $q$ and $k_{F}$ is the 
Fermi wave number, follows from the model we use for the el-ph 
coupling function $g^{2}(Q)=(g^{2}/Q_{c}^{2})\theta(Q_{c}-Q)$, 
where $Q=q/(2k_{F})$ and $\theta$ is the Heaviside function. 
The aim of this model is to simulate 
the effect of strong electronic correlations on the el-ph matrix element. 
In fact, according to different theoretical approaches, the tendency 
of the electronic correlation \cite{zeyher} and the weak screening \cite{weber}
is to suppress the scattering processes with large momentum transfer and, 
at the same time, to enhance the small 
$Q=q/(2k_{F})$ scatterings. 

\begin{figure}
\vbox to 6cm{\vfill
\centerline{\psfig{figure=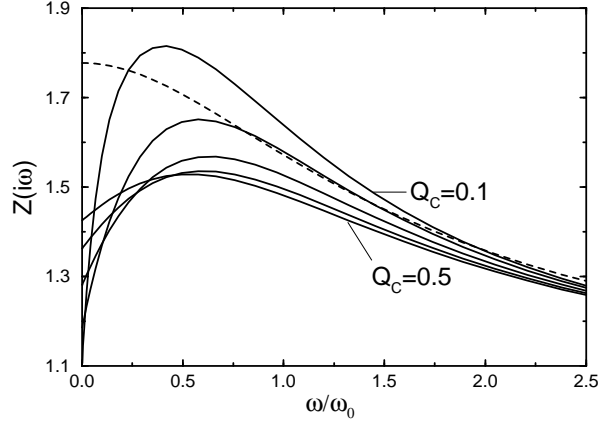,width=9cm}}\vfill}
\caption{Renormalization function $Z(i\omega)$
for $\omega_{0}/E_{F}= 0.2$ and $\lambda=1$. 
Solid lines: case with vertex correction for
$Q_{c}=0.1,\,0.2,\ldots,\,0.5$.
Dashed lines: case without vertex correction.}
\label{zetafun}
\end{figure}

The calculated $Z(i\omega)$ is shown in Fig. \ref{zetafun} 
as a function of $\omega/\omega_{0}$
for $\omega_{0}/E_{F}=0.2$, $\lambda=1.0$.
When the vertex correction is not included (dashed line), $Z(i\omega)$ 
takes the maximum value $Z_{max}$ at $\omega=0$. The deviation
from the adiabatic limit [$\lim_{\omega_{0}/E_{F}\rightarrow 0} Z_{max}=1+\lambda$] 
is due only to the finiteness of the electronic band \cite{grima2}.
When the vertex correction is included (solid lines) the maximum of $Z(i\omega)$ 
is shifted towards higher values of the frequency $\omega$. 
For fixed values of $\omega_{0}/E_{F}$
and $\lambda$, the position and the amplitude of $Z_{max}$ depend on the
cut-off parameter $Q_{c}$. Moreover, the value of $Z(i\omega)$
at $\omega=0$ is considerably lowered by the presence of the vertex correction. 
This feature can be understood by considering that at $\omega=0$ the electron-mass
renormalization factor $Z_{0}$ is mostly modified by the static limit 
of the vertex function, 
$P_{V}(i\omega_{n},i\omega_{m}\rightarrow i\omega_{n};Q_{c})$ ,
which is found to be negative \cite{grima2}. 
This can be recovered also by the following analytic expression valid for 
$\omega_{0}/E_{F}\rightarrow 0$ and $\omega_{0}/(Q_{c}^{2}E_{F})$ small:

\begin{equation}
\label{migdal3}
Z_{0}\simeq 1+\lambda-\lambda^{2}\frac{\pi}{4}\left(\frac{\omega_{0}}{Q_{c}^{2}E_{F}}\right) ,
\end{equation}
where the contribution of the negative static limit is given by the third term of the right hand side.
Note instead that for the superconducting transition
the situation is rather different, since the range of the relevant frequencies is
of order of $\omega_{0}$. In such a region of frequencies the vertex function
shows a complex behavior that can lead to an amplification or a suppression
of $T_{c}$ depending on the value of the parameter $Q_{c}$ \cite{grima1,grima2}.
This different role of the nonadiabatic contribution in $Z_{0}$ and $T_{c}$ 
reflects the fact that the inclusion of the vertex corrections
cannot be merely interpreted as a simple renormalization of the el-ph
coupling. 

\begin{figure}
\vbox to 6cm{\vfill
\centerline{\psfig{figure=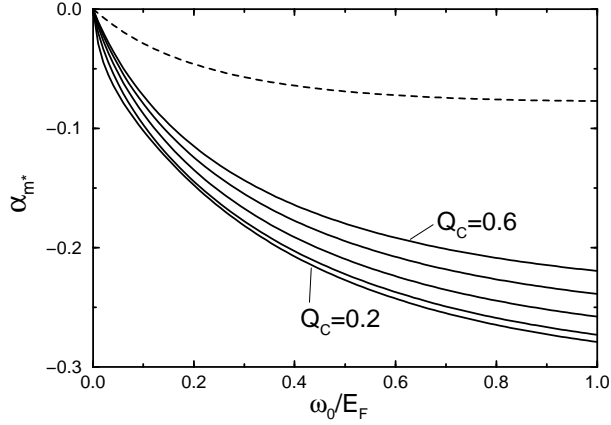,width=9cm}}\vfill}
\caption{Coefficient $\alpha_{m^{*}}$ of the isotope effect of the
effective electronic mass calculated for $\lambda=1$.
Solid lines: case with vertex correction for 
$Q_{c}=0.2,\,0.3,\ldots,\,0.6$.
Dashed lines: case without vertex correction.}
\label{alfavse}
\end{figure}

We show in Fig. \ref{alfavse} the calculated coefficient of  
$\alpha_{m^{*}}$ as a function of the adiabatic parameter 
$\omega_{0}/E_{F}$ for $\lambda=1$ and different values of $Q_{c}$.
$\alpha_{m^{*}}$ takes negative values regardless the absence (dashed line) or the
presence (solid lines) of the vertex correction. 
The presence of a nonzero $\alpha_{m^{*}}$ for the case without vertex correction has
to be ascribed only to finite-band effects, which give rather small absolute values
of the isotope coefficient. 
The inclusion of the vertex correction amplifies the nonadiabatic effects and leads 
to more negative values of $\alpha_{m^{*}}$ with respect to the case without 
vertex correction.
This can be also inferred from eq.(\ref{migdal3}) which gives:

\begin{equation}
\label{migdal4}
\alpha_{m^{*}}\simeq-\frac{1}{2}\frac{\lambda^{2}}{Z_{0}}\frac{\pi}{4}
\left(\frac{\omega_{0}}{Q_{c}^{2}E_{F}}\right) ,
\end{equation}
where the finite band effects have been neglected ($\omega_{0}/E_{F} \ll 1$) 
and the vertex correction is evaluated only  up to the linear term 
$\omega_{0}/(Q_{c}^{2}E_{F})$. 
Note that in Fig. \ref{alfavse} $\alpha_{m^{*}}$ 
does not show in the whole region of $\omega_{0}/E_{F}$ a crucial 
dependence on $Q_{c}$ as instead is observed for the 
superconducting transition temperature $T_{c}$ \cite{grima1,grima2}.
As a consequence, the isotope effects on $T_c$ and $m^*$ are not
proportional as in the bi-polaron theory of superconductivity and in 
principle it is possible to have a negligible $\alpha_{T_c}$ and at the same time
an appreciably nonzero $\alpha_{m^*}$.

The result reported in Fig. \ref{alfavse} have been obtained by considering a structureless electronic density of states (DOS)
and are therefore relevant for three dimensional systems like the fullerene 
and the BaBiO$_{3}$ compounds.
However, photoemission measurements 
have given evidence for flat electronic bands close to the Fermi level for a
large class of cuprates \cite{shen}, opening the possibility that van Hove 
singularities (vHs) in the DOS could lead to important effects on both the 
normal and the superconducting phases \cite{newns}.
As discussed in Ref.\cite{baff1}, the presence
of a vHs near the Fermi level
represents an intrinsic nonadiabatic situation, where Migdal's theorem is not valid
even if $\omega_{0}/E_{F}\ll 1$.
It is natural therefore to investigate the effect on $\alpha_{m^{*}}$ of the
vertex correction calculated by using a vHs in the DOS.
Here we anticipate some preliminary results valid for $\omega_0/E_F\ll 1$
obtained by using the following DOS:

\begin{equation}
\label{vhs1}
N(\epsilon)=-N_{0}\ln\left|\frac{2\epsilon}{E}\right|,
\end{equation}
where $-E/2\leq\epsilon\leq +E/2$ and $N_{0}=N/E$ corresponds to a
constant DOS with $N$ electronic states. In this model, the vHs is located
at the Fermi energy $E_{F}=E/2$.
By making use of eq.(\ref{vhs1}), we have found that
for $\lambda_0=1$ and $\omega_0/E_F=0.05$, the isotope coefficient
$\alpha_{m^{*}}$ is $-0.29$ ($-0.6$) for $Q_c=0.4$ ($0.2$).
Therefore, the combined effect of the vHs and the vertex correction 
leads to a magnification of the values of $\alpha_{m^{*}}$, in better 
agreement with the experimental ones.

In summary, we have discussed the effect of different el-ph
regimes on the effective electronic mass $m^{*}$. 
The main result is that the possibility of having nonzero 
values of $\alpha_{m^{*}}$ cannot be associated exclusively to 
the presence of polaronic charge carriers. 
We have shown, in fact, that a negative $m^*$-isotope coefficient 
can be obtained by taking
into account the first el-ph vertex correction beyond Migdal's limit. 
The latter result appear to be of particular interest in view of the
fact that in the nonadiabatic theory of superconductivity the
breakdown of Migdal's theorem can lead to a strong enhancement of $T_{c}$
and various other effects \cite{grima1,grima2,baff1}.

\stars{We thank S. Ciuchi for helpful discussions and R. Zeyher 
for bringing to our
attention the work of Zhao et al..
C. G. acknowledges the support of an I.N.F.M. PRA project.}

\end{document}
\bye